\documentclass[aps,twocolumn,prl,amssymb,superscriptaddress]{revtex4}
\usepackage{amsmath}
\usepackage{graphicx}
\usepackage{color}
\usepackage{setspace}
\usepackage{multirow}
\usepackage[]{titlesec}
\usepackage{enumerate}
\usepackage{array}

\def\dEF{\chi_{\substack{\scalebox{0.6}{F}}}(\Delta H)}
\def\dGFA{\chi_{\substack{\scalebox{0.6}{GFA}}}}
\def\dGFAx{\chi_{\substack{\scalebox{0.6}{GFA}}}\left(\{x\}\right)}

\definecolor{Gray}{gray}{0.9}
\definecolor{LightCyan}{rgb}{0.88,1,1}
\begin{document}
\title{\Large Spectral descriptors for bulk metallic glasses \\ based on the thermodynamics of competing crystalline phases}

\author{Eric Perim}\thanks{These authors have contributed equally to this manuscript.} 
\affiliation{Department of Mechanical Engineering and Materials Science and Center for Materials Genomics, Duke University, Durham, North Carolina 27708, USA}
\author{Dongwoo Lee}\thanks{These authors have contributed equally to this manuscript.} 
\affiliation{School of Engineering and Applied Sciences, Harvard University, Cambridge, MA 02138, USA}
\author{Yanhui Liu}\thanks{These authors have contributed equally to this manuscript.} 
\affiliation{Department of Mechanical Engineering and Materials Science, Yale University, New Haven, CT 06511, USA}
\author{Cormac Toher}
\affiliation{Department of Mechanical Engineering and Materials Science and Center for Materials Genomics, Duke University, Durham, North Carolina 27708, USA}
\author{Pan Gong}
\affiliation{Department of Mechanical Engineering and Materials Science, Yale University, New Haven, CT 06511, USA}
\author{Yanglin Li}
\affiliation{Department of Mechanical Engineering and Materials Science, Yale University, New Haven, CT 06511, USA}
\author{W. Neal Simmons}
\affiliation{Department of Mechanical Engineering and Materials Science and Center for Materials Genomics, Duke University, Durham, North Carolina 27708, USA}
\author{Ohad Levy}
\affiliation{Department of Mechanical Engineering and Materials Science and Center for Materials Genomics, Duke University, Durham, North Carolina 27708, USA}
\author{Joost J. Vlassak}
\affiliation{School of Engineering and Applied Sciences, Harvard University, Cambridge, MA 02138, USA}
\author{Jan Schroers}
\affiliation{Department of Mechanical Engineering and Materials Science, Yale University, New Haven, CT 06511, USA}
\author{Stefano Curtarolo}
\email[]{email: stefano@duke.edu}
\affiliation{Department of Mechanical Engineering and Materials Science and Center for Materials Genomics, Duke University, Durham, North Carolina 27708, USA}

\date{\today}

\begin{abstract}

Metallic glasses attract considerable interest due to their unique combination of superb properties and processability. 
Predicting their formation from known alloy parameters remains the major hindrance to the discovery of new systems. 
We propose a descriptor based on the heuristics that structural and energetic {\it ``confusion''} obstructs crystalline growth, 
and demonstrate its validity by experiments on two well-known glass forming alloy systems. 
We then develop a robust model for predicting glass formation ability based on the geometrical and energetic features of 
crystalline phases calculated {\it ab-initio} in the {\sf AFLOW}  framework. 
Our findings indicate that the formation of metallic glass phases could be much more common than currently thought,
 with more than $17\%$ of binary alloy systems potential glass formers. 
Our approach pinpoints favorable compositions and demonstrates that smart descriptors, 
based solely on alloy properties available in online repositories, offer the sought-after key for accelerated discovery of metallic glasses.

\end{abstract}

\maketitle 

\section{Introduction}

Understanding and predicting the formation of multicomponent bulk metallic glasses (BMG)
is crucial for fully leveraging their unique
combination of superb mechanical properties \cite{chen2015does} and
plastic-like processability \cite{schroers2006amorphous, Schroers_blow_molding_2011,kaltenboeck2016shaping} for potential
applications
\cite{Johnson_BMG_2009,Greer_metallic_glasses_review_2009,Schroers_Processing_BMG_2010,johnson2016quantifying}.
The process underlying the formation of BMGs is still to be fully understood.
It involves a multitude of topological fluctuations competing during
solidification across many length scales \cite{kelton1991crystal, egami2006formation, egami2011atomic,ding2014combinatorial}.
Long-range processes, required by the typical non-polymorphic nature of the crystallization, and
atomic-scale fluctuations, precursors of short-range ordered competing phases
\cite{Busch_kinetics_BMG_2007}, are all pitted against each other and against glass formation \cite{kelton1991crystal, kelton2010nucleation, kelton1998new}.
Simulations of amorphous phases have been attempted to disentangle the mechanism of glass formation
\cite{sheng2006atomic,zhang2015glass,zhang2015origin,lee2003criteria,cheng2008relationship,cheng2009atomic,wang2015asymmetric,schoenholz2016structural},
within reasonable system sizes, using classical and semi-empirical potentials.
Although they have been successful in investigating the influence of factors such as the atomic size and packing on the glass-forming
ability (GFA), questions about competing crystalline phases and the dynamics of the process still remain, especially considering all the approximations demanded for
performing long molecular dynamics simulations.
Furthermore, adopting {\it ab-initio} methods has been challenging \cite{nmatHT}:
even while the most relevant metastable crystalline phases can be calculated and sorted by their energies
\cite{curtarolo:calphad_2005_monster,monsterPGM,curtarolo:art67,curtarolo:art53,curtarolo:art63},
the zero-temperature formalism, lacking vibrational free energy \cite{curtarolo:art65}, and
the absence of an underlying lattice on which to build configurational thermodynamics \cite{deFontaine_ssp_1994,curtarolo:art49}
make the problem impervious to direct computational analysis.

\begin{figure}[h!]
\centerline{\includegraphics[width=0.45\textwidth]{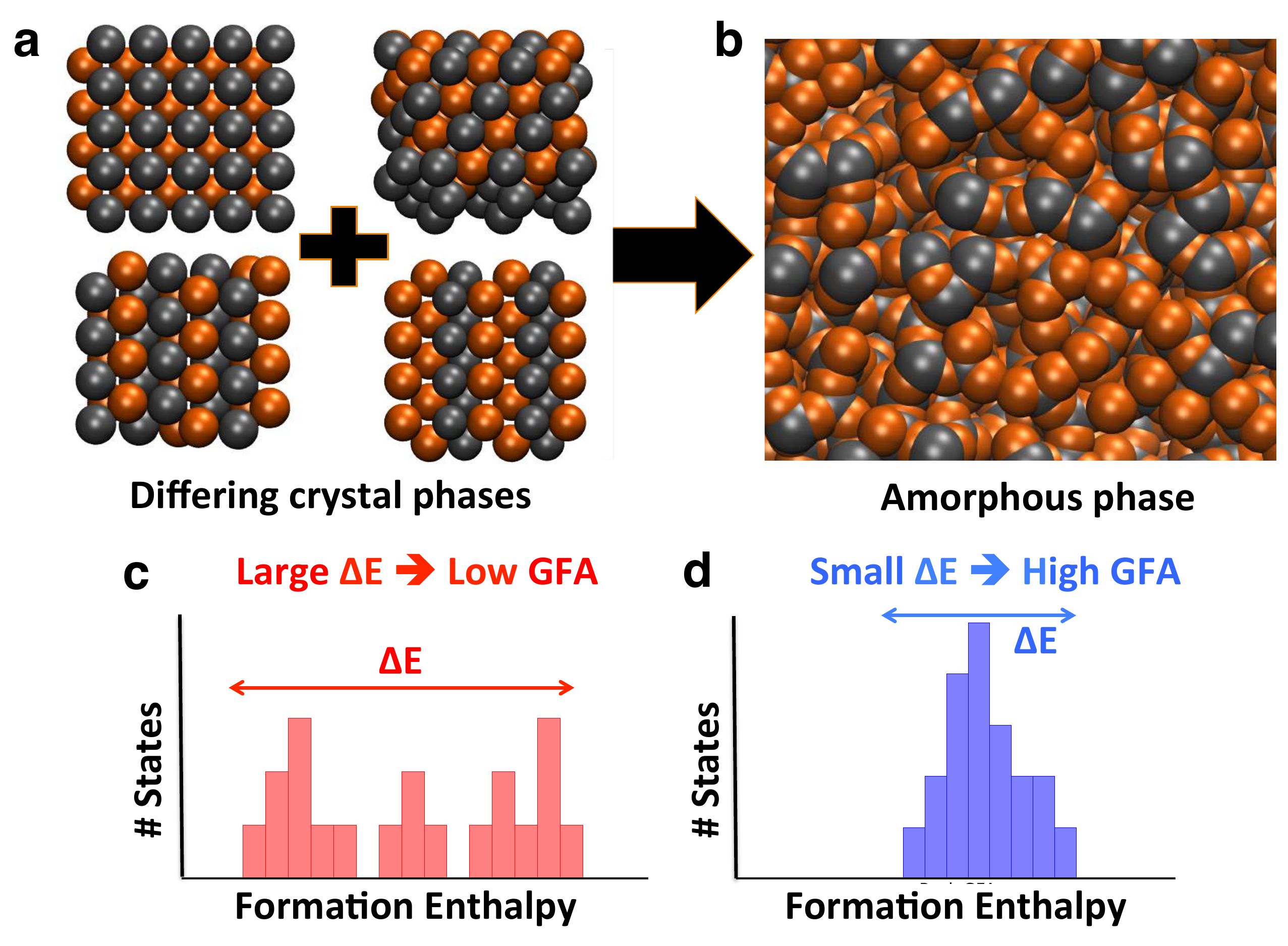}} 
\vspace{-2mm}
\caption{\small 
  {\bf Descriptor for confusion.} 
  If a particular alloy composition exhibits many structurally different stable and metastable crystal phases which have similar energies, 
  these phases will compete against each other during solidification, disrupting and frustrating the nucleation and crystallization processes, 
  ultimately leading to an amorphous structure. 
  {\bf (a)}  Distinct crystalline competing phases which may compete and lead to {\bf (b)} an amorphous structure.  
  Glass forming ability should be {\bf (c)} absent - {\bf (d)} present, when the thermodynamic density of states is low - high.} 
\label{fig1}
\end{figure}

\begin{figure*}[hbt]
\centerline{\includegraphics[width=0.80\textwidth]{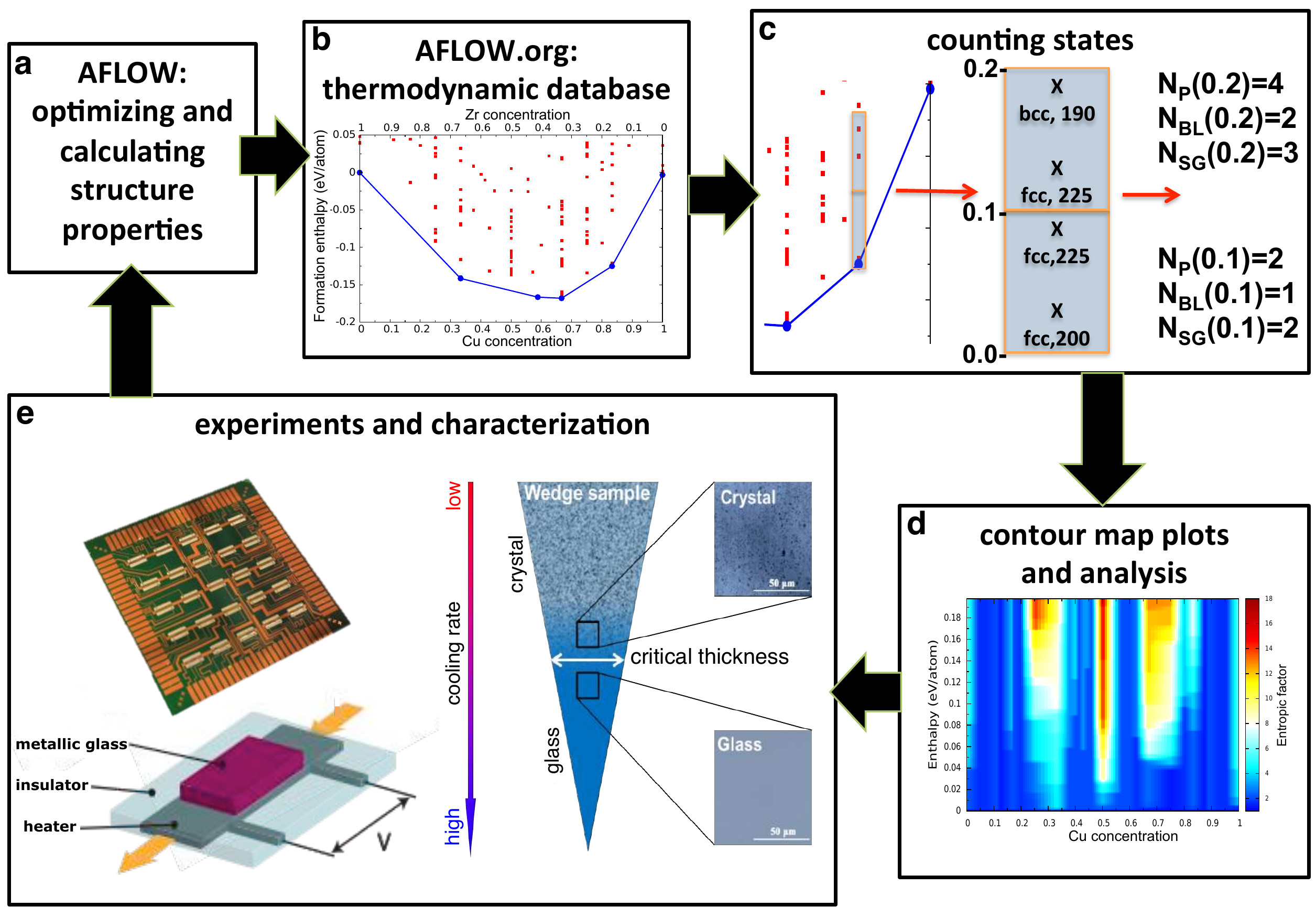}} 
\vspace{-2mm}
\caption{\small 
 {\bf Integration of experimental and computational approaches.}
   {\bf (a)} Multiple different structures for a given stoichiometry are built using the {\sf AFLOW} prototype libraries \cite{curtarolo:art65}, which are then optimized via VASP calculations under the {\sf AFLOW} standard settings \cite{curtarolo:art104}. 
   {\bf (b)} The resulting data is added to the open thermodynamic database {\sf AFLOW} \cite{curtarolo:art92,aflowlibPAPER}. 
   {\bf (c)} This data is accessed and used to obtain statistics on the cumulative distribution of entries ($N_P$), Bravais lattices types ($N_{BL}$) and space groups ($N_{SG}$) within a given formation enthalpy range (starting at zero). 
   {\bf (d)} Contour map plots are created from these distributions, allowing the identification of the best glass forming alloys.  {\bf (e)} Finally, experimental synthesis and characterization are used to verify the computational results.} 
\label{fig2} 
\end{figure*}

Descriptors for bulk glass formation --- correlations between the outcome (glass formation) and other material properties, possibly simpler to characterize \cite{nmatHT} ---
have been proposed based on structural \cite{miracle2004structural, wang2015asymmetric, cheng2009atomic, cheney2009evaluation}, 
thermodynamic \cite{cheney2009evaluation, cheney2007prediction, turnbull1969under, busch1995thermodynamics,lu2002new, johnson2016quantifying}, 
kinetic \cite{turnbull1969under,busch1998thermodynamics} and 
electronic structure considerations \cite{cheng2009atomic, alamgir2000electronic}. A few of these \cite{lu2002new, johnson2016quantifying}, have been considerably successful in correlating with the GFA. However, they rely on experimental data, such as the (reduced) glass transition temperatures, that can only be obtained once the glass has been synthesized, and, therefore, cannot be used to make predictions for systems that have not yet been experimentally studied. Consequently, a definite and clear picture for predicting GFA still remains to be found.

In a seminal paper \cite{greer1993confusion},
Greer speculated that ``confusion" during crystallization promotes glass formation.
However, challenges in a priori knowledge and abilty to quantify such confusion have left this direction mostly unexplored.
In this work we propose a  definition of this { ``confusion"} based on the following consideration.
During quenching, crystal growth will occur whenever fluctuations lead to the formation of a crystalline nucleus larger
than a critical size.
Therefore, in order to obtain an amorphous solid, the formation of critical size nuclei has to be hampered.
We postulate that the existence of multiple phases with very similar energy, implying similar probabilities of being formed,
but dissimilar structures, will lead to the formation of several distinct clusters
which will intimately compete and thus keep each other from reaching the critical size needed for crystallization.
To demonstrate the power of this ansatz, we first characterize confusion by the approximate thermodynamic density of distinct structural phases of metastable states,
obtained from \emph{ab-initio} calculations (Figure \ref{fig1}), and concurrent GFA measurements by combinatorial synthesis of alloy libraries and
high-throughput nanocalorimetry.
As test systems, we focus on CuZr and NiZr.
Among the known BMGs, CuZr is probably the most broadly studied \cite{liu2007cooling,xu2004bulk,wang2004bulk,mei2004binary}.
NiZr, on the other hand, is known for having poor GFA \cite{altounian1983crystallization,fukunaga2006voronoi}.
The contrast between the two glass formers, one strong and one weak, corroborates our ansatz. 
After having established the efficacy of our approach, we extend it into a robust numerical model for building GFA spectra. 
This extension establishes the strength of our approach, leading  to a descriptor that requires no experimental input and is computationally predictable, inexpensive and quick to calculate.

\begin{figure*}[htb!]
\centerline{\includegraphics[width=0.95\textwidth]{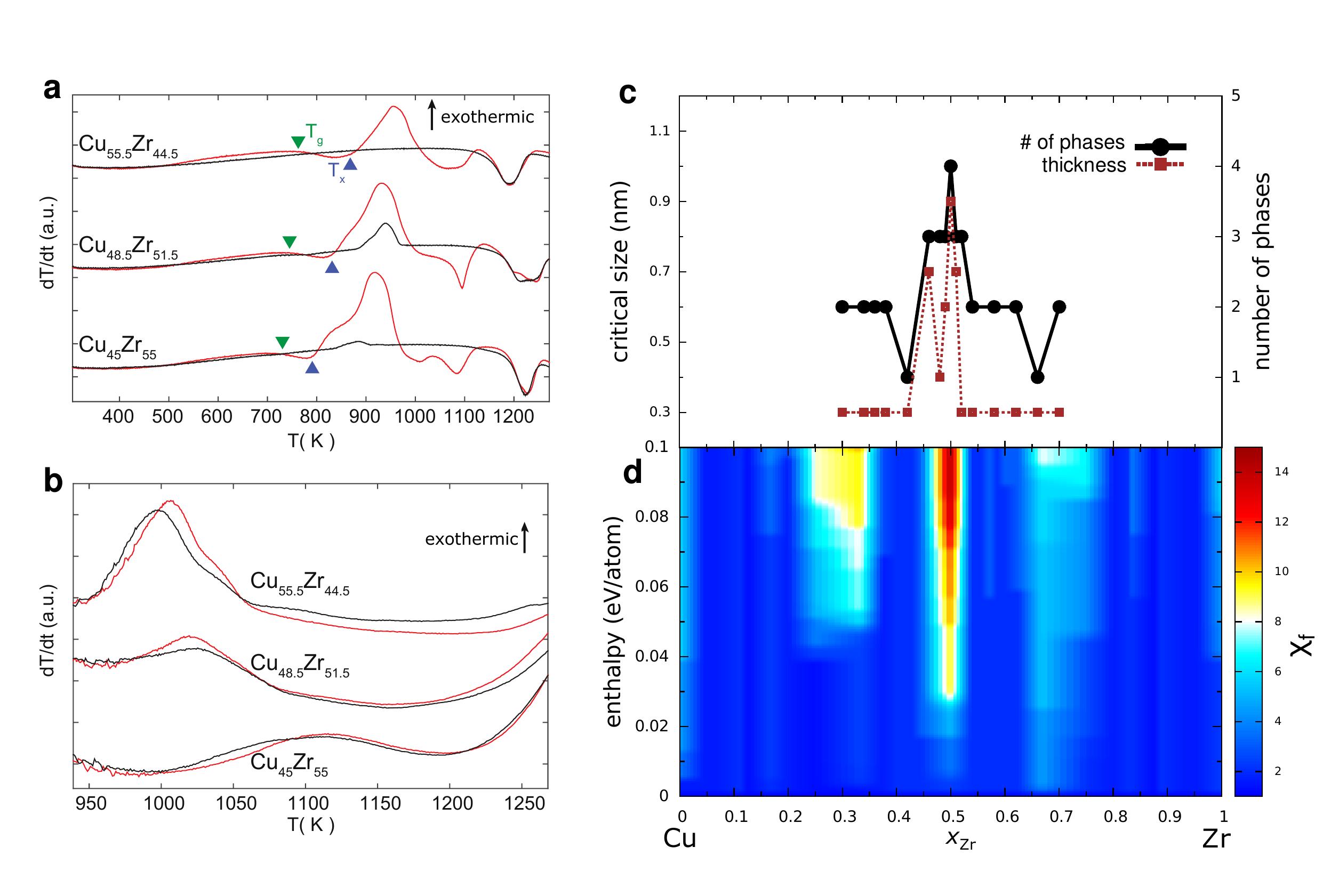}}
\vspace{-6mm}
\caption{\small 
 {\bf Experimental and theoretical analisys of CuZr.}
  {\bf (a)} Nanocalorimetry measurements during heating and {\bf (b)} cooling at different compositions. The first heating and cooling cycle measurements for the each composition are shown in red, and subsequent measurements are shown in black. 
  {\bf (c)}  Number of phases (solid black line) as measured using XRD, and thickness of the amorphous phase (dashed brown line), determined from the wedge shaped samples, as a function of composition. 
  {\bf (d)} Contour plot of the entropic factor as a function of formation enthalpy (zero corresponds to the ground state of the composition). 
  The color scale represents the entropic factor, calculated using Eq. (\ref{eq1}), for each composition and formation enthalpy difference. 
  This means that, for a given fixed composition ($x$-axis), all phases that are within a given formation enthalpy difference ($y$-axis) from the ground state of that specific composition are used to compute the entropic factor (color scale). 
  Note the sharp peaks both in the number of states observed in experiment and in the entropic factor at the Cu$_{50}$Zr$_{50}$ composition, indicating that the descriptor correctly identifies this composition as having the highest GFA.}
\label{fig3} 
\end{figure*}

\section{Results}
{\bf Using databases for materials discovery.} Carrying out electronic structure {\it ab-initio} calculations for the infinite number of available states for a given alloy system is obviously impossible, 
especially when no lattice model can be built \cite{deFontaine_ssp_1994,curtarolo:art49}, as in the case of BMGs.
Therefore, we adopt the agnostic approach of exploring structural prototypes mostly observed in nature for these types of systems.
The method, shown to be capable of reasonably sampling the phase space and predicting novel compounds \cite{curtarolo:calphad_2005_monster,monsterPGM,curtarolo:art67,curtarolo:art53,curtarolo:art49},
is expected to estimate the thermodynamic density of states of an alloy system. 
We use the binary alloy data available in the {\sf AFLOW} set of repositories \cite{curtarolo:art92,aflowlibPAPER} to count the number of different structural phases 
in a given formation enthalpy range as a function of the composition. 
This data was obtained utilizing the {\sf VASP} \cite{kresse_vasp, vasp, vasp_cms1996} code within the {\sf AFLOW} 
computational materials design framework \cite{curtarolo:art65,curtarolo:art104}, at the density functional theory level of approximation. 
The binary alloy systems are fully relaxed in accordance with the {\sf AFLOW} standard settings \cite{curtarolo:art104}; 
which uses the GGA-PBE \cite{PBE, PBE2} exchange-correlation, PAW potentials \cite{PAW, kresse_vasp_paw}, 
at least 6000 {\bf k}-points per reciprocal atom (KPPRA) and a plane wave cut-off at least 1.4 times the largest value recommended for the {\sf VASP} potentials of the constituents.
The multiple different crystalline phases for each particular stoichiometry are built from the {\sf AFLOW} library of common prototypes \cite{curtarolo:art65}.

{\bf A simple descriptor for glass formation.}
To quantify the level of disorder associated with an alloy system, we identify the most stable structures and
count all of the available phases at the corresponding compositions,
ordered by their formation enthalpy difference above the respective ground state, $\Delta H$.
This leads to a cumulative distribution of the number of phases, $N_P(\Delta H)$ (see Figure \ref{fig2}).
We also count the number of different Bravais lattice types, $N_{BL}(\Delta H)$, and space groups, $N_{SG}(\Delta H)$ among the phases in the distribution. 
These three quantities are combined into a single heuristic descriptor, called the { ``entropic factor'}', $\dEF$, 
defined as the cubic root of their product: 
\begin{equation} 
  \dEF=\sqrt[3]{N_P(\Delta H)\times N_{BL}(\Delta H)\times N_{SG}(\Delta H)}.
  \label{eq1} 
\end{equation}
$\dEF$ should be related to the configurational entropy at a given composition but, by taking into account the different
symmetries available to the system, it is more generally representative
of the frustration of the crystallization of a single homogeneous crystal structure.
Compositions with large $\dEF$ are expected to present structures with more disorder, thus leading to high GFA.
In this analysis, the formation enthalpies, Bravais lattices and space groups were determined
from the calculated energies and symmetries of the relaxed relevant structures.

X-ray diffraction (XRD) and scanning electron microscopy (SEM) measurements were performed on ingots of CuZr and NiZr alloys prepared by arc-melting the pure elements under an argon atmosphere. 
The alloys were re-melted and suction cast into a wedge-shaped cavity in a copper mold. 
The as-cast rods were cut into half along the longitudinal direction and polished to a mirror finish followed by etching. 
GFA was evaluated by observing the contrast change along the longitudinal direction under a scanning electron microscope . 
The critical thickness was determined at the transition from featureless contrast to a clearly observable microstructure, as shown in Figure \ref{fig2}. 
The crystalline and amorphous structures were further identified by XRD using a Cu-$K_\alpha$ source. 

We also synthesized and characterized thin-film samples deposited by magnetron sputtering elementary targets (99.99\% pure) inside a vacuum chamber with a base pressure better than $ 2\times10^{-7}$ Torr. Sputter deposition results in an effective quenching rate greater than $10^{9}$ K s$^{-1}$ \cite{liu2014structural}, allowing a broad range of alloys to be obtained in the amorphous state.

Nanocalorimetry measurements were performed on thin-film samples of the binary alloys using micromachined calorimetry sensors \cite{mccluskey2010combinatorial,mccluskey2010nano,lee2013scanning,lee2014low}. 
The measurements were performed in vacuum at nominal heating rates ranging from 2000 to 8500 K s$^{-1}$, and cooling rates of approximately 5000 K s$^{-1}$. 
All samples were repeatedly heated to 1300 K to evaluate the crystallization behavior both in the as-deposited state and after melt/quenching. 
Nanocalorimetry measurements reveal the glass transition, crystallization and liquidus temperatures. These quantities allow us to estimate GFA.

\begin{figure}[htb!] 
  \centerline{\includegraphics[width=0.49\textwidth]{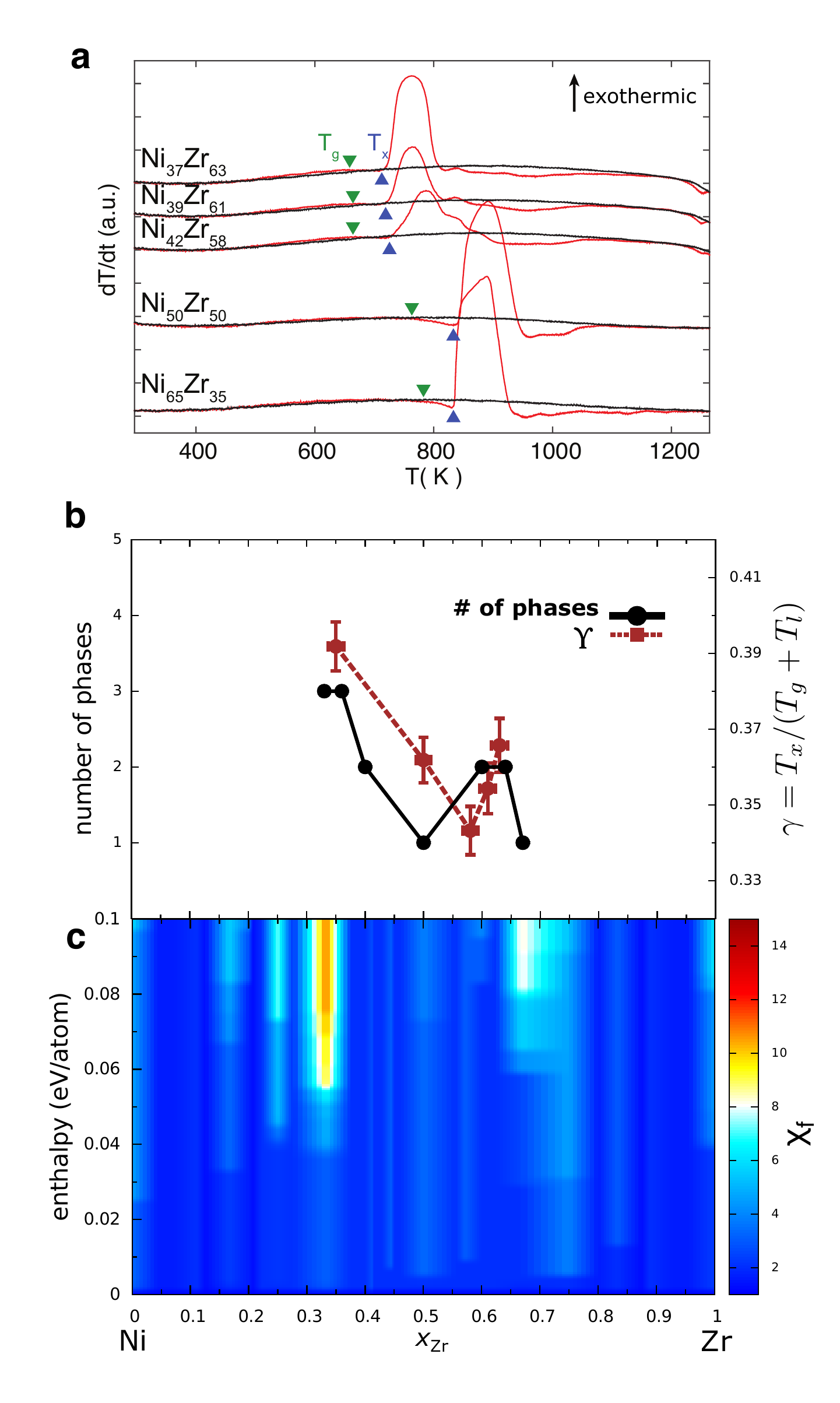}}
  \vspace{-6mm}
  \caption{\small 
    {\bf Experimental and theoretical analisys of NiZr.}
    {\bf (a)} Nanocalorimetry measurements for NiZr during heating at different compositions. 
    The first heating cycle measurements for each composition are shown in red, and subsequent measurements are shown in black.
    {\bf (b)} Number of phases (solid black line) as measured using XRD, and $\gamma$ descriptor calculated for NiZr alloys (dashed brown line). 
    {\bf (c)} Contour plot of the entropic factor as a function of formation enthalpy (zero corresponds to the ground state of the composition). 
    Note the sharp peaks both in the number of states observed in experiment and in the entropic factor at the Ni$_{35}$Zr$_{65}$ and Ni$_{65}$Zr$_{35}$ compositions.} 
  \label{fig4}
\end{figure}

Figures \ref{fig3}(a) and (b) show the nanocalorimetry results for the CuZr binary alloy with compositions in the bulk glass forming region. 
Each measurement consisted of two thermal cycles in which the thin-film samples were heated to above the melting point and then quenched. 
All samples show clear signals corresponding to glass transition, crystallization, and melting when first heated from the as-deposited state, indicating that they were deposited in the amorphous state, Figure \ref{fig3}(a).
A better glass former has a lower critical cooling rate, so the amount of amorphous phase recovered after melt/quenching should scale with GFA. 
We observe in Figure \ref{fig3}(a) that the magnitude of the crystallization peak after the first thermal cycle changes significantly with composition: Cu$_{48.5}$Zr$_{51.5}$ has the strongest crystallization peak 
and is thus expected to have the highest GFA among the samples tested; Cu$_{55.5}$Zr$_{44.5}$ on the other hand has no discernible crystallization peak. 
This result is confirmed by calorimetry measurements obtained after cooling from the melted state: 
the heat released upon solidification results in an exothermic peak in the cooling curve; 
the magnitude of this peak scales with the amount of crystalline phase formed on quenching and should be inversely proportional to the GFA, Figure \ref{fig3}(b). 
The experimentally observed number of phases and the amorphous phase thickness obtained from the casting experiments are shown in Figure \ref{fig3}(c). 
The calculated entropic factor, Figure \ref{fig3}(d), can be compared with these two quantities, and the results show very good agreement between all methods that Cu$_{50}$Zr$_{50}$ is the best glass forming composition.

Figure \ref{fig4} shows similar measurements for the NiZr alloy system, which has been shown to be a weak glass former \cite{altounian1983crystallization,fukunaga2006voronoi}. 
Although as-deposited samples were amorphous and showed distinct crystallization peaks, subsequent melt/quenching did not produce any amorphous samples,
and no crystallization peaks are observed in scans obtained after melting (Figure \ref{fig4}(a)). Instead, we use $\gamma\equiv{T_x}/{(T_g+T_l)}$, defined in Ref. \onlinecite{lu2002new} and shown in Figure \ref{fig4}(b), 
as a less direct measure of GFA. Figures \ref{fig4}(b) and (c) show strong correlation between the experimental measurements and the entropic factor descriptor.
There is a very weak GFA peak around Ni$_{35}$Zr$_{65}$ according to the experimental measurements, which is predicted to be around Ni$_{30}$Zr$_{70}$ by the entropic factor. 
A more pronounced peak is measured around Ni$_{65}$Zr$_{35}$, which is also successfully predicted in the same region by the entropic factor descriptor. 
Thus, the new proposed descriptor correlates well with the traditional empirical indicators of glass formation in metallic alloys, with an accuracy of the order of $5\%$ in  composition, which is quite satisfactory.
In addition, comparing Figures \ref{fig3}(c) and \ref{fig4}(c), it is clear that the entropic factor exhibited by the high GFA alloy CuZr is significantly higher
than that shown by the low GFA NiZr alloy, thus correctly pointing out the more favorable alloy system for glass formation.
These results validate our ansatz and show that crystalline phase data can be used in order to predict the formation of amorphous phases.

{\bf Descriptor for Glass Forming Ability.} 
Following this demonstration of the promise of our characterization of structural confusion, we proceed to enhance it into a broader and more quantitative model.
This requires several steps: the ansatz is that the presence of  highly dissimilar structures with very similar enthalpy correlates with GFA and the
descriptor should contain factors describing enthalpy proximity, structural similarity and appropriate normalizations.
Once the descriptor is defined, it will be confronted with experimental results and a threshold will be found self consistently. 
Finally the formalism will be applied to our online repository {\sf AFLOW} for appropriate statistical analysis and potential suggestions of glass-forming alloys.
 
{\bf Enthalpy proximity.}
The descriptor should favor states with enthalpy close to the ground state. 
This is captured by a Boltzmann factor:
\begin{equation}
\begin{aligned}
f(H_i)=&\exp\left (\frac{-\left | H_i-H_0 \right |}{k_BT_0}\right ) \times \\
& \times \begin{cases}
1, & \text{$H_i<0$} \\
e^{-H_i/k_BT_0}, & \text{$0\leqslant H_i<50 {\rm\, meV}$} \\
0, & \text{$50 {\rm\, meV}\leqslant H_i$},
\end{cases}
\label{enthalpy}
\end{aligned}
\end{equation}
in which $H_0$  is the lowest enthalpy for a given concentration, and $T_0$ is room temperature.
The inclusion of phases with positive formation enthalpy is necessary due to glass formation occurring at higher temperatures, at which higher enthalpy phases become accessible \cite{widom2015high}.
The cut-off value for including positive formation enthalpy phases is taken to be 50 meV $\sim 600K$,  of the same order as the glass transition temperature of several metallic glasses.

{\bf Structure similarity.}
To correlate properties of structures having different decorations of the underlying lattice, we use a lattice-free formalism,
the expansion in local \underline{A}tomic \underline{E}nvironments  (AE) \cite{villars:factors}. 
The AE of an atom is defined as the polyhedron formed by the atoms present in the neighborhood 
up to the distance of the maximum gap in the radial distribution function.
A given structure has the corresponding AE calculated for each and every unique atom and then is expanded as:
\begin{equation}
\begin{aligned}
 \left | \psi \right \rangle = \sum c_i \left | AE_i \right \rangle, \ \ \ \  \left \langle AE_i | AE_j \right \rangle = \delta_{ij}, \\ 
c_i= \left \langle AE_i | \psi \right \rangle,  \ \ \ \ \sum c_i^2=1,
\label{expansion}
\end{aligned}
\end{equation}
where $\psi$  is a vector representing a given atomic structure. In this representation, the scalar product
 \begin{equation}
\left \langle \psi | \psi' \right \rangle = \sum\limits_{ij} \left \langle AE_i \right | c_i^{*}c'_j \left | AE_j \right \rangle = \sum\limits_{i} c_i^{*}c'_i
 \label{scalarProduct}
 \end{equation}
  is used to quantify the structural (dis)similarity between two distinct structures.
The structural similarity factor is taken as an exponential having the maximum when 
$\left\langle\psi_i |\psi_0\right\rangle\!=\!0$ (structures are dissimilar) and decaying to 0 at $\left\langle\psi_i |\psi_0\right\rangle\!=\!1$ (structures are similar):
\begin{equation}
\begin{aligned}
g(\left |\psi_i\right \rangle)=&\exp\left (\frac{-\theta}{|1-\left \langle \psi_i |\psi_0\right \rangle |}+\theta|1-\left \langle \psi_i |\psi_0\right \rangle |\right )\times \\
& \times \left (1-\overline{\left \langle \psi_i |\psi_j\right \rangle} \right )^2,
\label{morphology}
\end{aligned}
\end{equation}
where $\theta\!=\!0.25$ is a constant, based on an analysis of the available experimental data and kept constant for the entire study.
The multiplicative coefficient is added to take into account the limitation that the exponential is 
taken with respect to the lowest enthalpy state at a given concentration $\psi_0$, 
and therefore structural similarity among metastable states needs to be accounted for
by taking the average scalar product between metastable structures $i$ and $j$, $\overline{\left \langle \psi_i |\psi_j\right \rangle}$, 
computed over all possible combinations for a given stoichiometry $\{x\}$.

{\bf Normalization.}
The normalization is represented by this expression computed for each stoichiometry $\{x\}$ of a given alloy system:
\begin{equation}
h\left(\{x\}\right)=\mathrm{\# \:of\: entries\: within\: cutoff\: at\: stoich.\ \{x\}}.
\label{normalization}
\end{equation}

{\bf Glass Forming Ability descriptor.}
Combining  Eqns. (\ref{enthalpy}), (\ref{morphology}) and (\ref{normalization}) we generate the GFA descriptor
evaluated by summing through structures $i$ at a fixed stoichiometry $\{x\}$:
\begin{equation}
  \dGFAx=\frac{\sum_i f(H_i) g(\left | \psi_i \right \rangle)}{h\left(\{x\}\right)}.
  \label{dGFA}
\end{equation}
A large peak of $\dGFAx$ is expected to indicate good glass forming ability at a particular concentration $\{x\}$.

\begin{figure*}[htb!]
\begin{centering}
\centerline{\includegraphics[width=\textwidth]{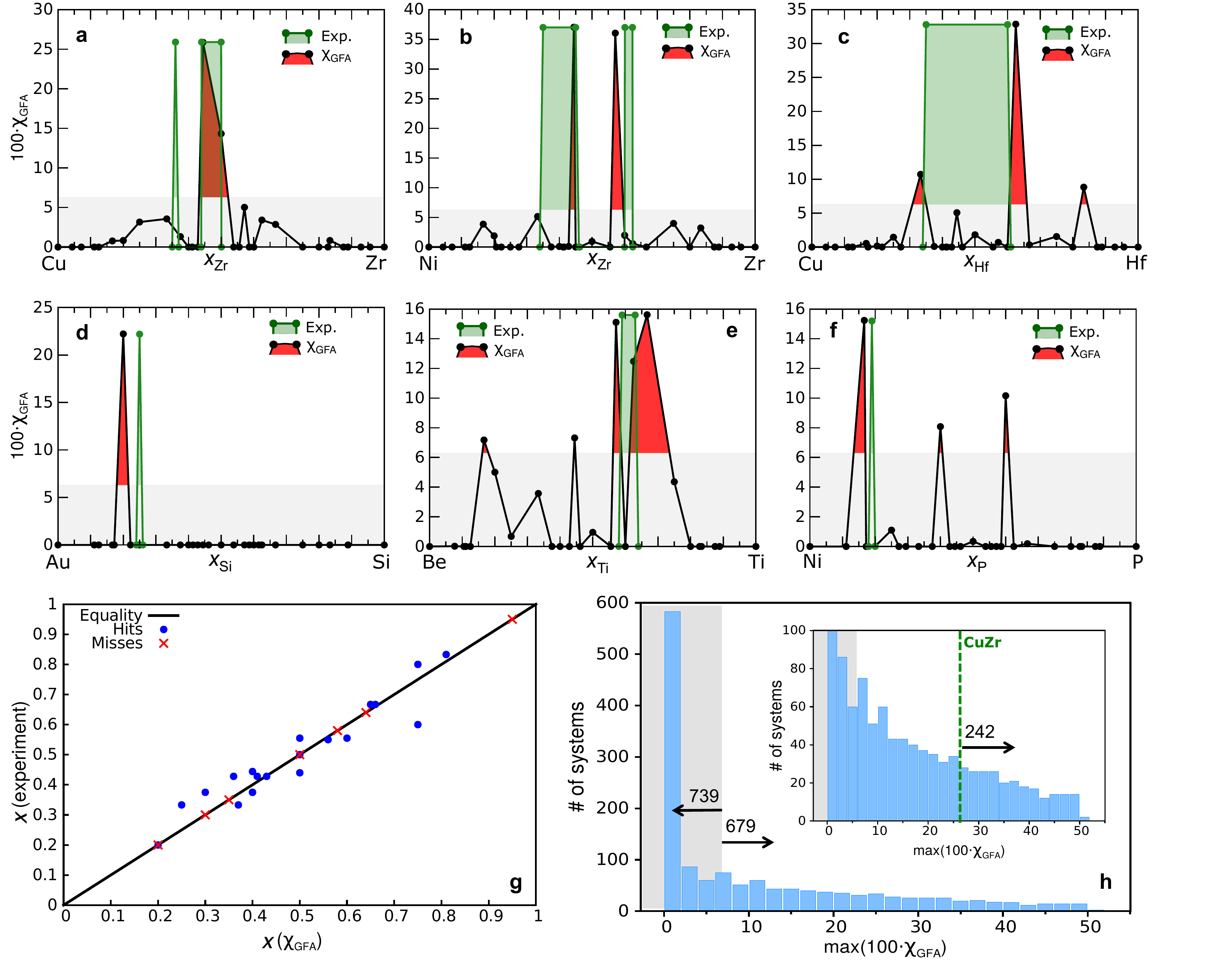}}   
\caption{\small
  {\bf Glass forming ability descriptor spectra for different alloys.}
  Predictions are shown in black line/solid red fill, experimentally reported compositions are shown in green line/transparent green fill and the area under the threshold is shown in grey. 
  {\bf (a)} CuZr (reported glass formers Cu$_{50}$Zr$_{50}$, Cu$_{56}$Zr$_{44}$ and Cu$_{64}$Zr$_{36}$ \cite{liu2007cooling,li2008matching}); 
  {\bf (b)} NiZr (reported glass formers Ni$_{1-x}$Zr$_{x}$ with $0.35\!<\!x\!<\!0.45$ and $0.60\!<\!x\!<\!0.63$ \cite{altounian1983crystallization}); 
  {\bf (c)} CuHf (reported glass former Cu$_{1-x}$Hf$_{x}$ with $0.35\!<\!x\!<\!0.60$ \cite{inoue2004formation}); 
  {\bf (d)} AuSi (reported glass former Au$_{75}$Si$_{25}$ \cite{klement1960non}); 
  {\bf (e)} BeTi (reported glass former Be$_{1-x}$Ti$_{x}$ with $0.59\!<\!x\!<\!0.63$ \cite{tanner1979metallic}); 
  {\bf (f)} NiP (reported glass former Ni$_{81}$P$_{19}$ \cite{calvo1997structural}).
  {\bf (g)} Reported versus predicted glass forming concentrations for the 16 training systems.
  Missed glass formers are noted as red crosses. 
  {\bf (h)} Statistical distribution of the maximum peak GFA value for 1418 different binary alloys. Inset shows a close up of the same plot. Area under the threshold is shown in grey.}
\label{fig5}
\end{centering}
\end{figure*} 

{\bf Comparison with experiments and threshold value.}
The GFA descriptor $\dGFAx$ was trained with respect to the available experimental data on binary metallic glasses. 
This data is scarce and sparse. 
Usually, only glass forming compositions are reported
\cite{liu2007cooling,altounian1983crystallization,inoue2004formation,klement1960non,tanner1979metallic,calvo1997structural,logan1975structure,graczyk1980atomic,grzeta1985hydrogen,yu2006fabrication,leonhardt1999solidification,wang2010ni,ruppersberg1980observation,lewis1976stabilities,chen1981structure,altounian1985crystallization},
hindering the training of the descriptor to determine true-negatives. 
Equipped with these 16 systems' comparisons, we search for a threshold which is found self-consistently as the 
lowest value maximizing the ratio ``peak hits versus misses'' without increasing the number of false positives.
The threshold is found to be $\sim0.063$. Figures \ref{fig5} (a)-(f) show six binary examples comparing the predicted glass-forming compositions versus known experimental ones (arbitrarily assigned the highest descriptor value obtained for each corresponding system).
The systems are
CuZr \cite{liu2007cooling,mei2004binary,wang2004bulk,xu2004bulk,li2008matching},
NiZr \cite{altounian1983crystallization},
CuHf \cite{inoue2004formation,xia2006glass,figueroa2013metallic},
AuSi \cite{klement1960non}, 
BeTi \cite{tanner1979metallic}, and 
NiP \cite{calvo1997structural}.
When $\dGFA>0.063$ we claim the existence of a glassy phase. 
As mentioned earlier, CuZr is probably the most studied binary metallic glass, due to its high GFA and accessible constituent materials. 
Figure \ref{fig5} (a) compares our prediction with the experimentally reported glass forming compositions of Cu$_{50}$Zr$_{50}$, Cu$_{56}$Zr$_{44}$ and Cu$_{64}$Zr$_{36}$ \cite{liu2007cooling,li2008matching,mei2004binary,wang2004bulk,xu2004bulk}, showing good agreement. 
For the CuHf alloy system, a glass forming range is reported in Cu$_{1-x}$Hf$_{x}$ between $0.35<x<0.60$ \cite{inoue2004formation,xia2006glass,figueroa2013metallic}. 
As shown in Figure \ref{fig5}(c), we only register peaks at the extremes of this range, possibly suggesting a two glass coexistence in that composition range.
Overall, of the 16 systems we analyze, 15 are correctly identified as glass-formers with our descriptor (reliability $\sim\!94\%$). 
However not all of the peaks are reproduced. Out of the 26 peaks available in the 16 systems, 19 are found (reliability  $\sim\!73\%$). 
Qualitatively the predicted concentrations are always close to the experimental values
but due to the finite set of compositions spanned and the limited number of structures at each composition in our {\sf AFLOW} repository, they are not strictly accurate.
Figure \ref{fig5}(g) shows the correlation between predicted and reported concentrations, 
which is quite good, with a root mean square deviation of $5.4\%$ for the successfully predicted ones (the {\sf AFLOW} database has 200-250 different optimized structures for each of these systems.
Several concentrations are computationally challenging to parameterize, hindering a uniform sampling of the spectrum).
Table \ref{GFAtable} lists the systems and compositions used for the development of $\dGFA$.

\begin{center}
\begin{table}[h!]
\begin{tabular}{ l | m{1.9cm} | l }  
 Reported & Predicted & Ref. \\
 \hline 
 \hline 
Cu$_{50}$Zr$_{50}$, Cu$_{56}$Zr$_{44}$ and Cu$_{64}$Zr$_{36}$ & Cu$_{100-n}$Zr$_{n}$,  $50\!\!<\!\!n\!\!<\!\!55$ & \cite{liu2007cooling,li2008matching} \\  
 \hline 
Ni$_{100-n}$Zr$_n$,  $35\!\!<\!\!n\!\!<\!\!45$, $60\!\!<\!\!n\!\!<\!\!63$  & Ni$_{42.8}$Zr$_{57.2}$ Ni$_{55.5}$Zr$_{44.5}$ & \cite{altounian1983crystallization} \\ 
 \hline 
Cu$_{100-n}$Hf$_n$,   $35\!\!<\!\!n\!\!<\!\!60$ \ & Cu$_{16.7}$Hf$_{83.3}$ Cu$_{37.5}$Hf$_{62.5}$ Cu$_{66.7}$Hf$_{33.3}$ & \cite{inoue2004formation} \\  
 \hline 
Au$_{75}$Si$_{25}$  & Au$_{80}$Si$_{20}$ & \cite{klement1960non} \\ 
 \hline 
Be$_{100-n}$Ti$_n$,   $59\!\!<\!\!n\!\!<\!\!63$ & Be$_{33.3}$Ti$_{66.7}$ Be$_{42.8}$Ti$_{57.2}$ & \cite{tanner1979metallic} \\ 
 \hline 
Ni$_{81}$P$_{19}$ & Ni$_{60}$P$_{40}$ Ni$_{40}$P$_{60}$ Ni$_{83.3}$P$_{16.7}$ & \cite{calvo1997structural} \\ 
 \hline 
Au$_{20}$La$_{80}$ & Au$_{20}$La$_{80}$  Au$_{37.5}$La$_{62.5}$  Au$_{62.5}$La$_{37.5}$   Au$_{80}$La$_{20}$ & \cite{logan1975structure}  \\  
 \hline 
Au$_{35}$Ni$_{65}$ & - & \cite{graczyk1980atomic}  \\ 
 \hline 
Be$_{100-n}$Zr$_n$,   $50\!\!<\!\!n\!\!<\!\!70$ & Be$_{37.5}$Zr$_{62.5}$ & \cite{tanner1979metallic}  \\ 
 \hline 
Cu$_{50}$Ti$_{50}$,  Cu$_{58}$Ti$_{42}$,  Cu$_{66}$Ti$_{34}$ & Cu$_{37.5}$Ti$_{62.5}$,  Cu$_{66.7}$Ti$_{33.3}$ & \cite{grzeta1985hydrogen}  \\ 
 \hline 
Nb$_{30}$Ni$_{70}$,  Nb$_{40.5}$Ni$_{59.5}$ &  Nb$_{44.4}$Ni$_{55.6}$   Nb$_{50}$Ni$_{50}$   Nb$_{62.5}$Ni$_{37.5}$   Nb$_{83}$Ni$_{17}$ & \cite{yu2006fabrication,leonhardt1999solidification}  \\ 
\hline
Ni$_{60}$Ta$_{40}$ & Ni$_{55.6}$Ta$_{44.4}$ & \cite{wang2010ni}  \\ 
 \hline 
Ni$_{40}$Ti$_{60}$ & Ni$_{16.7}$Ti$_{83.3}$  Ni$_{25}$Ti$_{75}$  Ni$_{37.5}$Ti$_{62.5}$  Ni$_{50}$Ti$_{50}$  Ni$_{55}$Ti$_{45}$  Ni$_{66.7}$Ti$_{33.3}$    & \cite{ruppersberg1980observation}  \\ 
 \hline 
Pd$_{100-n}$Si$_n$,   $5\!\!<\!\!n\!\!<\!\!25$ & Pd$_{60}$Si$_{40}$ & \cite{lewis1976stabilities} \\  
 \hline 
P$_{25}$Pt$_{75}$ & P$_{20}$Pt$_{80}$ P$_{33.3}$Pt$_{66.7}$ P$_{44}$Pt$_{56}$ & \cite{chen1981structure}  \\ 
 \hline 
Fe$_{100-n}$Zr$_n$,   $57\!\!<\!\!n\!\!<\!\!80$ & Fe$_{42.8}$Zr$_{57.2}$ & \cite{altounian1985crystallization}  \\ 
\end{tabular}
 \caption{\small
   {\bf Experimentally reported glass formers.}
   List of 16 reported glass forming alloys used for training the spectral descriptor. Whenever a broad glass forming region was reported we 
   counted two peaks, one at the beginning of the region and one at the end. This approach leads to a total of 26 peaks used as references. 
   An empty entry at the second column means that no glass-forming composition was predicted, {\it i.e.}, a miss. 
   The second column includes both peaks that correspond to the reported ones as well as a few
   that do not correspond to any of the reported glass forming alloys.}
 \label{GFAtable}
\end{table}
\end{center}

\section{Discussion}
{\bf Looking for new glass formers.}
The {\sf AFLOW} repository, 
containing a total of 1418 binary systems characterized by more then 330,000+ appropriate structural entries, was screened using the new descriptor.
The calculated $\dGFAx$ spectra for all of the binaries are summarized in Figure \ref{fig5}(h).
{\it In brevis}, the histogram of the maxima of $\dGFA$ shows that most of the systems, $\sim 52\%$ (739 over 1418), are below the threshold and therefore expected to be non-glass formers.
However, there are still many, $\sim 48\%$ (679), above the threshold and therefore potential glass-forming systems.
In particular $\sim 17\%$  (242) have $max(\dGFA)$ higher than the known good glass-former CuZr, and hence are highly plausible candidates for metallic glass-formers. 
The magnitude and sharpness of $\dGFA$ lead us to suggest the systems listed in Table \ref{PredictTable} for further experimental validation. The predicted GFA spectra for the suggested systems can be found in Supplementary Figs 1 and 2.

\begin{center}
\begin{table}[h!]
\begin{tabular}{ m{5.5cm} }
 Glass forming compositions \\
 \hline 
 \hline 
 Al$_{37.5}$La$_{62.5}$\\   \hline 
 Al$_{60}$Re$_{40}$\\   \hline 
 As$_{44.4}$Nb$_{55.6}$; As$_{60}$Nb$_{40}$ \\   \hline 
 Co$_{33}$Zn$_{67}$\\   \hline 
 As$_{20}$Pd$_{80}$; As$_{62.5}$Pd$_{37.5}$\\  \hline 
 Ba$_{83.3}$Zn$_{16.7}$\\  \hline 
 Be$_{55}$V$_{45}$\\   \hline 
 Bi$_{60}$Pt$_{40}$\\  \hline 
 Cr$_{44.4}$Rh$_{55.6}$ \\  \hline 
 Fe$_{37.5}$Nb$_{62.5}$ \\  \hline 
 Fe$_{40}$P$_{60}$; Fe$_{62.5}$P$_{37.5}$\\ \hline
 Ga$_{40}$Ir$_{60}$ \\  \hline
 Ge$_{62.5}$Rh$_{37.5}$ \\  \hline 
 Hf$_{44.4}$Pd$_{55.6}$ \\  \hline 
 Hf$_{55.5}$Re$_{44.5}$; Hf$_{60}$Re$_{40}$\\  \hline 
 La$_{60}$Pb$_{40}$ \\  \hline
 La$_{60}$Pd$_{40}$ \\  \hline
 Mg$_{40}$Pb$_{60}$ \\  \hline 
 Mn$_{62.5}$Si$_{37.5}$ \\  \hline
 Nb$_{55.5}$Os$_{44.5}$ \\  \hline
 Nb$_{37.5}$Si$_{62.5}$ \\  \hline
 P$_{83.3}$Pd$_{16.7}$ \\  \hline
 Pb$_{62.5}$Sc$_{37.5}$;  Pb$_{80}$Sc$_{20}$ \\  \hline
 Pd$_{44.4}$Zn$_{55.6}$;  Pd$_{60}$Zn$_{40}$ \\  \hline
 Pd$_{37.5}$Zr$_{62.5}$;  Pd$_{55.5}$Zr$_{44.5}$ \\
\end{tabular}
\caption{\small
  {\bf Potential candidate glass formers.}
  List of unreported compositions that are predicted to present high glass forming ability (spectra are shown in the Supplemental Figs. 1 and 2).}
\label{PredictTable}
\end{table}
\end{center}

Overall, our analysis implies that the existence of metallic glass phases 
could be a very common phenomenon in nature and that the missed experimental observations would be mostly due to the difficulty in achieving the appropriate
quenching rates and/or in the choice of compositions.
For addressing the latter problem, the rational interrogation of online repositories through carefully-trained heuristic descriptors that capture the physical essence of the problem could become the long sought quantum leap in the field.

We propose a novel predictor for metallic glass formation that is based on the structural and thermodynamic properties of competing crystalline phases which we calculate from first principles. This predictor stems from the concept that competition between energetically-similar crystalline phases frustrates crystallization and thus promotes glass formation. It was developed into a robust numerical descriptor using formation enthalpies and structural similarity measures based on atomic environments. Detailed nanocalorimetry experiments verify the validity of this approach for two model systems, CuZr and NiZr. The non-reliance on experimental data allows for the construction of GFA spectra for 1418 different binary alloy systems,  by leveraging extensive libraries of computed crystalline phase data such as {\sf AFLOW}. Our results predict that $17\%$ of binary alloy systems are capable of glassifying, including many whose synthesis has not been previously reported in the literature, suggesting that there is great uncharted potential for new discoveries in this field. 

\section{Acknowledgements}
This work was supported by the National Science Foundation under DMREF Grants No. DMR-1436151, DMR-1436268, and DMR-1435820.
Experiments were performed in part at the Center for Nanoscale Systems at Harvard University (supported by NSF under Award No. ECS-0335765), 
at the Materials Research Science and Engineering Center at Harvard University (supported by NSF under Award No. DMR-1420570), and
at the Materials Research Science and Engineering Center at Yale University (supported by NSF under Award No. DMR-1119826).
Calculations were performed at the Duke University - Center for Materials Genomics.
E.P., C.T. and S.C. acknowledge partial support by the DOD-ONR (N00014-13-1-0635, N00014-14-1-0526).

\section{Author contribution}
S.C. proposed the entropy and spectral descriptors;
E. P. wrote the code under the supervision of C. T. and O. L.;
D.L. performed the nanocalorimetry experiments under the supervision of J. V.;
Y.L., P.G. and Y.L. executed the combinatorial synthesis and XRD and SEM measurements under the supervision of J. S..
All authors contributed to the discussion and to the writing of the manuscript.

\section{Methods}
{\small
{\bf Sample preparation.}
The ingots of CuZr and NiZr alloys were prepared by arc-melting the pure elements under an argon atmosphere. 
The alloys were re-melted and suction cast into a wedge-shaped cavity in a copper mold. 
The as-cast rods were cut into half along the longitudinal direction and polished to a mirror finish followed by etching. \\
We also synthesized thin-film samples deposited by magnetron sputtering elementary targets (99.99\% pure) inside a vacuum chamber with a base pressure better than $ 2\times10^{-7}$ Torr. Sputter deposition results in an effective quenching rate greater than $10^{9}$ K s$^{-1}$ \cite{liu2014structural}, allowing a broad range of alloys to be obtained in the amorphous state. \\
{\bf Nanocalorimetry experiments.}
Nanocalorimetry measurements were performed on thin-film samples of the binary alloys using micromachined calorimetry sensors \cite{mccluskey2010combinatorial,mccluskey2010nano,lee2013scanning,lee2014low}. 
The measurements were performed in vacuum at nominal heating rates ranging from 2000 to 8500 K s$^{-1}$, and cooling rates of approximately 5000 K s$^{-1}$. 
All samples were repeatedly heated to 1300 K to evaluate the crystallization behavior both in the as-deposited state and after melt/quenching. \\
{\bf First-principle calculations.}
All DFT calculations were carried in accordance with the {\sf AFLOW} standard settings, which are described in detail in Ref. \onlinecite{curtarolo:art104}. \\
{\bf Calculation of AEs.}
To discern AEs we generate $N \times N \times N$ supercells for the structures under consideration, $N$ being odd and larger than or equal to 3. 
All distances are calculated with respect to the atoms in the central cell and only within a sphere centered on each atom with a radius chosen so as to guarantee that it is always enclosed by the supercell.
If less than one hundred neighbors are contained within this sphere then the supercell size is increased in order to meet this requirement.
This is done in order to guarantee sufficient sampling as well as to avoid spurious gaps around the edges of the supercell.
Exceptions to this rule are considered when either there are two or more gaps of similar size, or when the AE defined by this rule generates convex polyhedra in which atoms are contained on the faces (instead of exclusively in the vertices). 
For the first case, two gaps are considered equivalent if they differ by 0.05\AA\ or less. In this case we adopt the gap which defines the smaller AE \cite{daams_villars:environments_2000}. 
For the second case, whenever atoms are detected within a surface, the AE is reconstructed using the largest gap which defines an AE smaller than the initial one. After generating an AE, each of its vertices (atoms) are classified by the number and type of different faces (either triangular or quadrilateral) meeting at that point. Finally, an AE is described in terms of the number of each type of vertex \cite{daams:cubic_environments}.
 It should be emphasized that, using this classification, slight distortions on the AEs are completely ignored, and thus we account only for significant differences in crystal structures.}

\section{Data availability}
All the {\it ab-initio} alloy data is freely available to the public as
part of the AFLOW online repository and can be accessed through
{\sf www.aflow.org} following the REST-API interface \cite{curtarolo:art92}.

\end{document}